# Magnetic Field Tunable Capacitive Dielectric:Ionic-liquid Sandwich Composites

Ye Wu

The University of Texas at San Antonio, Department of Electrical and Computer Engineering, One UTSA Circle, San Antonio, TX 78249.

**Abstract**: We examined the tunability of the capacitance for $GaFeO_3$-ionic liquid-$GaFeO_3$ composite material by external magnetic and electric field. Up to 1.6 folds of capacitance tunability could be achieved at 957kHz with voltage 4v and magnetic field 0.02T applied. We show that the capacitance enhancement is due to the polarization coupling between dielectric layer and ionic liquid layer.

**1.INTRODUCTION**

There could be two effective strategies for strengthening the electro-magnetic coupling in materials. One strategy is the introduction of magnetic ion dopant to make nonmagnetic electric materials magnetic. This is shown in the synthesis of room temperature ferromagnetic or antiferromagnetic semiconductors,[1]-[5] which is made by dopping magnetic particles into III-V semiconductors. Due to the low solubility of magnetic particles in the semiconductor composites, the electro-magnetic effects is relatively small.[1]

Another strategy is to reconstruct electric counterparts of conventional magnetic compound. This involves the reconstruction of the charge density and thus the electrical modification of the electric polarization in the magnetic material system. One approach could be integrating guest material which contains high carriers concentration. One candidate for this approach could be the ionic liquid, a conductive fluid which has been intensively used for carriers engineering of strongly correlated electron system. For example, it has been used for introducing novel superconducting states, achieving electroresistance tunability. [6]-[14]. In this letter, we sandwiched the ionic liquid between two $GaFeO_3$ layers, a well known multiferroic materials.[15]-[21]

We performed the study of dielectric response of the $GaFeO_3$-ionic liquid-$GaFeO_3$ composite material under external magnetic field and bias voltage. We expect this composite material could hold the electro-magneto-dielectric properties. The motivation for this strategy is to find out the highly tunable dielectric response of composite material with respect to the external field, which is crucial for possible future applications in tunable microwave devices.

**2.Experiments.**

The $GaFeO_3$ thick film samples(110nm) used in the study were produced by polymer assisted deposition method.[22] The $GaFeO_3$ phases were confirmed by x-ray powder diffraction. The phases was confirmed by XRD (Figure 1(a)). The structure of $GaFeO_3$ was refined and drawn by GSAS software package. [23][24] Then the refined unit cell is drawn by DRAWxtl.[25] One drop of ionic liquid 1-n-Butyl-3-methylimidazo-li-umhexafluorophosphate (Alfa Aesar) was placed between two slides of $GaFeO_3$ and mechanically pressed to have a thickness around $100\,\mu m$. This gave us the $GaFeO_3$-ionic liquid-$GaFeO_3$ sandwich structure. The electric resistivity was measured by a standard 4 probes measurement setup equipped with an oven. The temperature hadn't been increased to higher than 398K for sample damage concern. The magnetic field strength up to 21A/m was generated by a DC magnet. A dc power source( Agilent E3648A) was used for applying external bias voltage. We measured the capacitance using a LCR meter (Hp HEWLETT PACKARD 4284A).In order to eliminate possible leakage current from power source, the sample was placed between two commercial capacitors which provide isolation. The measurement unit was further electrically shielded and guarded in order to prevent leakage current. The magnetic field direction is perpendicular to the sample surface. The capacitance of the GaFeO$_3$-ionic liquid-GaFeO$_3$ sandwich structure was measured with respect to the external magnetic field and under dc voltages.

**3.Theory.**

For the following, it would be important to explore the nature of magento-electro-dielectric coupling in this composite material. Since it contains the bulky ionic liquid layer and the magnetic layer, we may suppose the total electric polarization is the superposition of the two kinds of polarization generated from these two distinct layers. This assumption is based on our dielectric measurement of ionic liquid and GaFeO$_3$ material. The dielectric response of either GaFeO$_3$ or ionic liquid doesn't show any distinct characterization when electric field or magnetic field is applied.

We suppose there are three major kinds of electric polarization in this composite material. The first is the polarization P$_1$ that is generated from the dielectric layer of GaFeO$_3$. The second is P$_2$ from the ionic liquid layer and the third is from the interface polarization between the GaFeO$_3$ and ionic liquid layer, which could be represented by $P_{12} = CP_1 + \varepsilon_0 E_1 = CP_2 + \varepsilon_0 E_2$. (3.1)

Here C represents the coupling factor[26] between the GaFeO$_3$ and ionic liquid layer. E$_1$ and E$_2$ are the local electric field which could be fitted from the experimental data. The total polarization is

$$P = \lambda_1 P_1 + \lambda_2 P_2 + \lambda_3,$$ (3.2)

where $\lambda_1, \lambda_2$ and $\lambda_3$ are the constants which could be fitted from the experimental data. P$_{12}$ is not shown in this expression, since it could be also represented by P$_1$ or P$_2$ in Eq.(3.1).

P$_1$ could be calculated using classical method,[27] $P_1 = \dfrac{4\pi(E - C_\omega \sum C_\beta H^\beta)}{\varepsilon_0}$ (3.3)

where C$_\omega$ is the parameter which depends on the electric signal frequency. C$_\beta$ is the constant that is associated with $\Sigma H^\beta$. E is the electric field across the sample.

In the following, we will calculate $P_2$. This calculation of the second kind of polarization could be started from the analysis of ions dynamic with respect to the magnetic field.

The ionic liquid would involve with the process of the exchange of two ions, which is actually the replacement of an anion by a cation. [28][29]We consider an anion in position A is replaced by a cation in position B. For simplicity, the velocity of cation from A to B is supposed to be the same with that of anion from B to A. The charge $Q$ would undergo a magnetic force $QHv$ and travel with a certain distance of $x$.

The work required in this exchange is

$$W = (-Q)(\varphi_A - \varphi_B) + Q(\varphi_B - \varphi_A) + 2QHv_{AB}x_{AB} = 2Q(\varphi_B - \varphi_A) + 2QHvx, \quad (3.4)$$

Where $2QHvx$ is the work attributed to the magnetic force.

Using Boltzmann's law, the equilibrium ionic concentrations at the two spots are in the ratio of

$$\frac{c_B}{c_A} = \exp(\frac{-W}{kT}) = \exp[\frac{2Q(\varphi_A - \varphi_B) - 2QHvx}{kT}] \quad (3.5)$$

We assume the total ionic concentration is constant, which is $c_B + c_A = c$. $\quad (3.6)$

Using Eq.(3.5) and Eq.(3.6), we can derive $c_A = \dfrac{c}{1+\exp(t)}$, (3.7) $c_B = \dfrac{c\exp(t)}{1+\exp(t)}$, (3.8) where

$$t = \frac{2Q(\varphi_A - \varphi_B) - 2QHvx}{kT} = \frac{2Q\varphi(x) - 2QHvx}{kT}. \quad (3.9)$$

The Poisson's law indicates $\dfrac{d^2\varphi(x)}{dx^2} = \dfrac{-\rho(x)}{\varepsilon}$. $\quad (3.10)$

Here the charge density could be expressed as $\rho(x) = F(c_B - c_A) = \dfrac{F(\exp(t)-1)c}{1+\exp(t)}$. $\quad (3.11)$

$F$ is the Faraday constant.

Plugging Eq.(3.9) and Eq.(3.11) into Eq.(3.10), we can generate $\dfrac{d^2 t}{dx^2} = \dfrac{-Fc}{\varepsilon}\dfrac{\exp(t)-1}{\exp(t)+1}$, $\quad (3.12)$ which could be solved as $\frac{dt}{dx} = \sqrt{\frac{Fc}{2\varepsilon}[t - 2\ln(1+e^t)]}$ . (3.13)

We suppose the charge on the dielectric surface is q, then the charge in the diffuse layer would be -q when the electroneutrality is concerned. It should be noted that

$$-q = \int_x^\infty \rho(l)dl .$$ (3.14)

Combining Eq.(3.11) and Eq.(3.14), we can get $q = \sqrt{2Fc\varepsilon[t - 2\ln(1+e^t)]}$ (3.15)

Using Eq.(3.13), we can have

$$\frac{dt}{dx} = \sqrt{\frac{Fc}{2\varepsilon}[t - 2\ln(1+e^t)]} = \frac{q}{2\varepsilon}$$ (3.16)

Then the polarization could be calculated as

$$P_2 = \gamma \frac{\partial q}{\partial \varphi} = \gamma \frac{\partial q}{\partial t}\frac{\partial t}{\partial \varphi} = \frac{\gamma\sqrt{2Fc\varepsilon}Q(1-e^t)}{kT\sqrt{t - 2\ln(1+e^t)}(1+e^t)} = \frac{\gamma\sqrt{2Fc\varepsilon}Q(1-e^t)}{kT\sqrt{t}(1+e^t)} .$$ (3.17)

where $\gamma = dE_0/A$ is the constant that associated with sample dimension $d$, sample area $A$ and localized field $E_0$.

Combining Eq.(3.3) and Eq.(3.17), the total capacitance of dielectric: ionic structure is calculated as

$$C_{tot} = P/\gamma = \frac{4\pi\lambda_1(E - C_\omega\sum C_\beta H^\beta)}{\varepsilon_0\gamma} + \frac{\lambda_2\sqrt{2Fc\varepsilon}Q(1-e^t)}{kT\sqrt{t}(1+e^t)} .$$ (3.19)

## 4. Results and analysis.

Fig.A1(a) (listed in Appendix A) shows the XRD result of GaFeO$_3$ film. We used GSAS package to refine the structure. The lattice length was refined as, a=8.751Å, b=9.398Å, c=5.079Å. The lattice angles are α=β=γ=90°. The space group is Pc21n. As shown in Fig.A1(b)(Appendix A), every gallium atom is located at the center of octahedron formed by six ozone atoms. We measure the resistance dependence on temperature. When 295K<T<388K, this GaFeO$_3$:ionic liquid sandwich structure shows quasi-semiconductor transport (Fig.A1(c)).

It should be noted that the electrochemical window of the ionic liquid is in the range not higher than than 6v. For this reason, the external voltage applied was only 4v,5v and 6v, respectively, which could make physical and chemical properties of the ionic liquid stable.

In order to investigate the electro-magneto-dielectric properties of this composite material, we investigated the capacitance tunabilities corresponding to varied electric frequencies, voltages and magnetic field (Fig.A2). The capacitance tunabilities in Fig.A2 are defined as $\Delta C/C_0$, where $C_0$ is the origin capacitance with magnetic field H=0.0069T. $\Delta C$ is the variation of capacitance corresponding to $C_0$ . It shows that the capacitance tunability with voltage 4v is relatively higher than those with voltage 0v, 5v and 6v. This result indicates the 4v may be the optimized bias voltage. 160% capacitance tunability could be achieved at the frequency 957kHz with the voltage 4v and the magnetic field 0.02T.

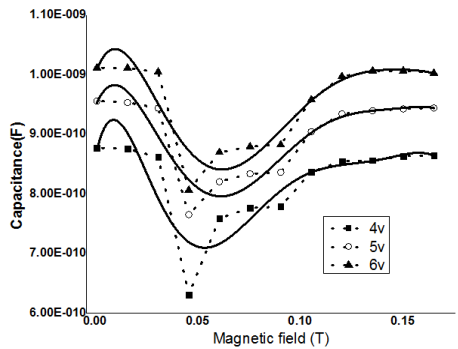

Figure 1 Comparing the experimental and simulated data of magnetocapacitance for the 957k Hz and with varied voltage and magnetic field applied.( The dash lines represent the experimental results while the solid lines represent the simulated results)

Figure 1 compares the experimental and simulated data of the capacitance with fixed voltage of 4v,5v and 6v and with varied magnetic field. The order of the capacitance dependence on the magnetic field β(Eq.(3.19))was fitted as 7, which suggested a very strong coupling between the magnetic field and polarization.

It should be noted that the experimental data corresponding to 957Hz in fig.1 was actually picked up from one of the data points in fig.A2(b). It would be cumbersome if we have to show all the simulated data corresponding to different frequency. For this reason we only choose the presentation of the experimental and simulated data corresponding to one specific frequency of 957Hz. The other results with other frequency are similar, which all show nonlinear dependence of the capacitance on the magnetic field. In closing, we do show a very precise quantitative analysis of our data in terms of theoretical calculations capturing all the details of the composite material investigated in the present scope; we also show the trend of transition and predict the trend of the capacitance dependence on the external magnetic field and voltage. This simple model simulates the important characteristics of capacitance with varied magnetic field and electric field in this composite material.

## 4.CONCLUSION.

$GaFeO_3$-ionic liquid-$GaFeO_3$ composite material shows quasi-semiconductor transport when 295K<T<388K. We examined the electro-magneto-dielectric effect on this structure. We find that up to 1.6 folds capacitance tunability could be achieved at 975kHz with the interaction of the magnetic field 0.02T and bias voltage 4v. We present a mechanism which takes account of the coupling between the ionic liquid layer and the solid dielectric layer. The effect shows a possible route to use the ionic liquid integrated with the dielectric material and suggests a simple way to introduce tunable dielectric permittivity.

## ACKNOWLEDGMENTS

The authors thank Prof.Robert Whetten in UTSA Physics Department for many useful discussions. This work was supported by the National Science Foundation under the Grants ECCS #1002380 and DMR #0844081.

**Appendix A**:

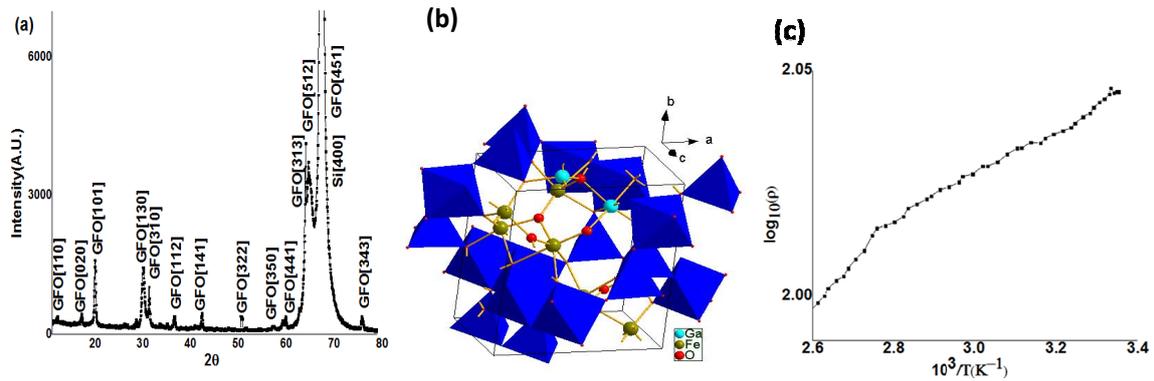

Figure A1.(a)XRD profile of GaFeO$_3$.(b)the unit cell of GaFeO$_3$.(c)Resistance vs. temperature characterization of the GaFeO$_3$: ionic liquid sandwich structure.

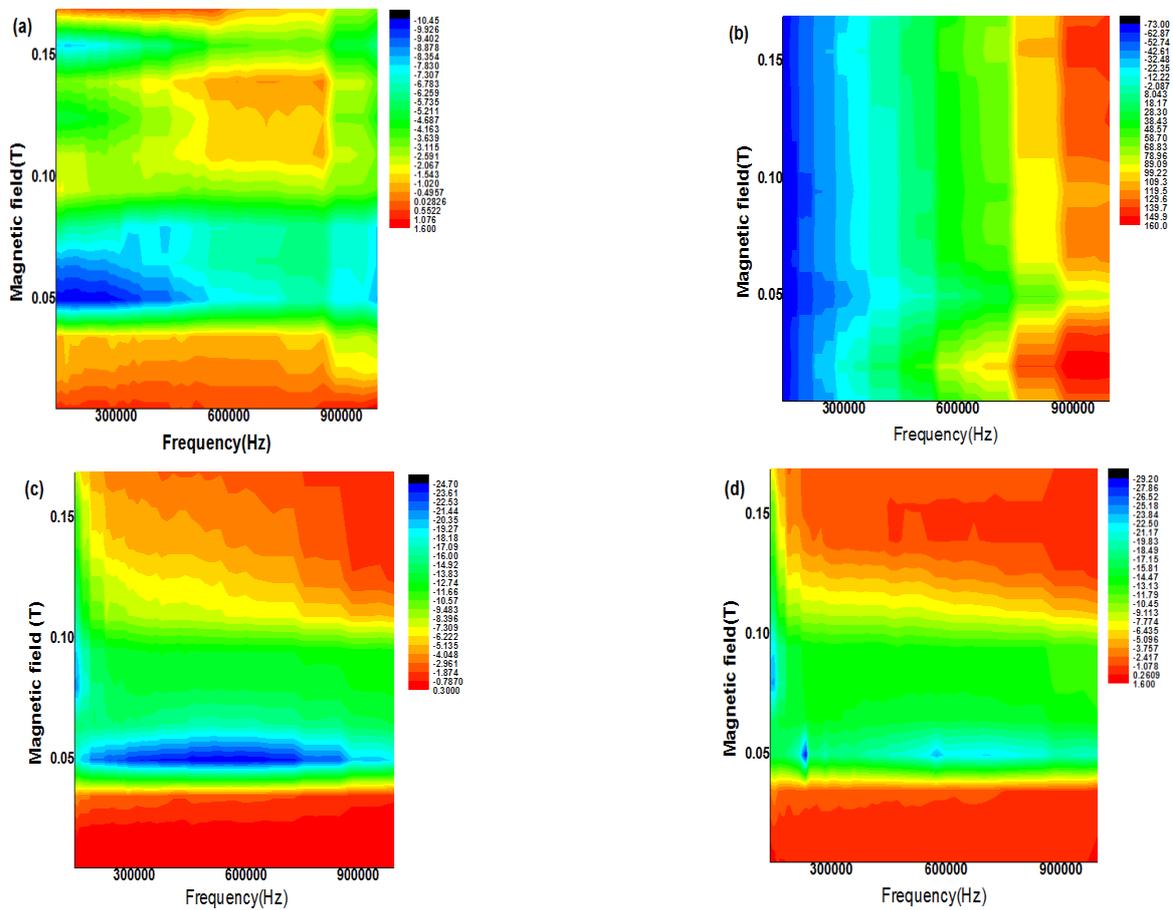

Figure A 2 . 2D capacitance tunability mapping corresponding to different bias voltage (a)0v (b) 4v (c)5v(d)6v.